\documentclass{aa}
\usepackage{graphicx}

\begin{document}

\title{FUSE observations of hot white dwarfs}

\author{B. Wolff\inst{1}
        \and
        J.W. Kruk\inst{2}
        \and
        D. Koester\inst{1}
        \and
        N.F. Allard\inst{3,4}
        \and
        R. Ferlet\inst{3}
        \and
        A. Vidal-Madjar\inst{3}
       }

\offprints{B. Wolff}

\institute{Institut f\"ur Theoretische Physik und Astrophysik,
           Universit\"at Kiel, D-24098 Kiel, Germany\\
           \email{wolff, koester@astrophysik.uni-kiel.de}
           \and
           Center for Astrophysical Sciences,
           The Johns Hopkins University,
           Baltimore, MD, 21218, USA\\
           \email{kruk@pha.jhu.edu}
           \and
           CNRS Institut d'Astrophysique de Paris,
           98bis Boulevard Arago,
           F-75014 Paris, France\\
           \email{allard, ferlet, alfred@iap.fr}
           \and
           Observatoire de Paris-Meudon,
           D\'{e}partement Atomes et Mol\'{e}cules en Astrophysique,\\
           F-92195 Meudon Principal Cedex
          }

\date{Received 6 April 2001 / Accepted 27 April 2001}

\abstract{We have analyzed FUSE observations of six hot white dwarf stars:
four DA white dwarfs with $T_{\rm eff} \ga 45\,000$~K, the DAO
\object{Feige\,55} ($T_{\rm eff} \approx 55\,000$~K), and the DA
\object{CD\,$-38^{\circ}$\,10980} ($T_{\rm eff} \approx 24\,000$~K).
Photospheric lines from Si\,IV, P\,V, and S\,VI
can be observed in the majority of the five hotter objects. Feige\,55 shows
also several other heavier elements. The measured abundances agree only partly
with the predictions of the radiative levitation theory.
We attribute this to current limitations of the models and the probable
presence of mass loss. In the spectrum of CD\,$-38^{\circ}$\,10980, we have
observed the quasi-molecular satellites of L$\beta$. This confirms
theoretical predictions about the visibility range for these features.
         }

\maketitle

\sloppy


\section{Introduction}

The ultraviolet part of the electromagnetic spectrum is of special interest
for investigations of hot white dwarf stars because of strong lines from
hydrogen, helium, and several heavier elements. Our knowledge of the
chemical composition of white dwarf atmospheres depends strongly on
observations in this spectral region.
However, the ultraviolet is still not explored fully. At wavelengths
longer than L$\alpha$, intensive studies could be realized mainly with 
the International Ultraviolet Explorer (IUE) and
the Hubble Space Telescope (HST). At shorter wavelengths,
only very few observations of white dwarfs obtained with the
Voyager spectrometers, the Hopkins Ultraviolet Telescope (HUT), and the
Orbiting and Retrievable Far and Extreme Ultraviolet Spectrometer (ORFEUS)
exist until now.

The Far Ultraviolet Spectroscopic Explorer (FUSE)  provides for the first time
the possibility to investigate
the far ultraviolet at wavelengths shorter than L$\alpha$ for
a significant number of white dwarfs. FUSE is a NASA astronomy satellite
aiming at high resolution spectroscopy from 905 to 1187~\AA.
It has been developed in cooperation with the Canadian Space Agency and the
Centre National d'Etudes Spatiales of France. An overview of the FUSE mission
has been given by Moos et al. (\cite{Moos}).

In this paper, we analyze FUSE spectra of six white dwarfs with hydrogen-rich
atmospheres: four DAs with $T_{\rm eff} \ga 45\,000$~K,
the DAO Feige\,55 ($T_{\rm eff} \approx 55\,000$~K), and the cooler
DA CD\,$-38^{\circ}$\,10980 ($T_{\rm eff} \approx 24\,000$~K).
These objects were
observed within a larger project investigating the local interstellar medium.
We concentrate here on the photospheric lines to gain an improved
knowledge of the atmospheric composition of the five hotter white
dwarfs. CD\,$-38^{\circ}$\,10980 is interesting because of the
possibility to study the L$\beta$ satellite features.

\section{Observations and data reduction}

The observations used in this work are listed in Table~\ref{LOG}.
WD\,2309+105 (GD\,246) and WD\,2331$-$475 were observed through the
LWRS aperture
(30$\arcsec$ square); the other stars were all observed through the
MDRS aperture (4$\arcsec \times 20\arcsec$ slit).
The channel alignment is satisfactory for the LWRS observations, with
stable fluxes being obtained in all channels (see Sahnow et al. \cite{Sahnow}
for a discussion of the alignment issues for the four FUSE telescopes).
The MDRS observations, however, are affected by channel misalignments.  These
observations were generally obtained fairly early in the mission, before
the channel alignment behavior was well-characterized and while MDRS observing
procedures were still being developed.  As a consequence, the effective
exposure time for the LiF2, SiC1, and SiC2 channels can be lower than
that shown in Table~\ref{LOG}. 
The LiF1 channel is essentially always aligned, with
flux variations between exposures usually being less than 1\%.
The LiF2 channel alignment is fairly good in most cases, providing
50\% to 100\%
of the exposure time (except that detector 2 was off for P1041101).
The alignment of the SiC channels is more variable.  The effective exposure
time for the SiC channels ranges from zero (for WD\,0455$-$282 and the
first observation of WD\,1631+781) to nearly 100\%.  

\begin{table}[t]
\caption[]{Log of observations}
\begin{flushleft}
\begin{tabular}{l@{}c@{}lllr}
\noalign{\smallskip}
\hline
\noalign{\smallskip}
\multicolumn{3}{l}{WD Number} &Data set &Date &\multicolumn{1}{l}{Exp./s} \\
\noalign{\smallskip}
\hline
\noalign{\smallskip}
0455 &$-$ &282 &P1041101 &2000-02-03 &19\,668 \\
     &    &    &P1041102 &2000-02-04 &10\,121 \\
     &    &    &P1041103 &2000-02-07 &17\,677 \\
1202 &+   &608 &P1042101 &2000-02-26 &13\,763 \\
     &    &    &P1042105 &1999-12-29 &19\,638 \\
1620 &$-$ &391 &Q1100101 &2000-07-18 &4\,830 \\
1631 &+   &781 &P1042901 &2000-01-18 &22\,300 \\
     &    &    &P1042902 &2001-01-31 &30\,160 \\
2309 &+   &105 &P1044101 &2000-07-19 &14\,828 \\
2331 &$-$ &475 &P1044201 &2000-06-23 &19\,356 \\
\noalign{\smallskip}
\hline
\noalign{\smallskip}
\label{LOG}
\end{tabular}
\end{flushleft}
\end{table}

The raw data were processed using the standard calibration pipeline
(CALFUSE version 1.8.7), which produces a calibrated spectrum for each exposure
in each channel.
The relative alignment of the channels would often
change from one exposure to the next, causing the star to be placed in 
different locations in the spectrograph entrance aperture.  The resulting
shifts in the wavelength scale were negligible for the MDRS observations.
For the LWRS observations, the processed spectra for each exposure were
shifted manually as needed to correct for this effect.  
In addition, the WD\,1202+608 (Feige\,55) spectra were shifted
to coalign the photospheric 
absorption lines in order to correct for the orbital motion of this
binary system. The spectra for each channel were averaged following
coalignment of the individual exposures.

In most cases an absorption line would appear in two to four channels;
comparison of the spectra from different channels thus served
to verify that none of the lines being fit were affected by instrumental 
artifacts. In order to increase the signal-to-noise ratio for interesting
parts of the spectra, the observations for each channel
were resampled onto a common wavelength scale (with 0.006~\AA\ resolution)
and combined. If the effective exposure times varied significantly then
the individual spectra were weighted with the apparent flux level.

\section{Photospheric composition}
\subsection{Hot DA white dwarfs}
\subsubsection{Overview}

The chemical composition of white dwarf atmospheres is determined
mainly by the high surface gravity which leads to a chemical stratification
where the lightest element present, either hydrogen or helium, floats on top
of the atmosphere. For selective ions of heavier elements, the effect of
gravity can be compensated
if radiative acceleration is strong enough to prevent these ions
from sinking downwards. Theoretical calculations and observations in the
ultraviolet, extreme ultraviolet, and X-ray spectral regions have shown
that radiative levitation can support ions from
elements like carbon, nitrogen, oxygen, silicon, iron, nickel, and others
in the atmospheres of DA white dwarfs at temperatures higher than
$T_{\rm eff} \approx 40\,000$--50\,000~K
(see e.g. Barstow et al. \cite{Barstow93}, \cite{Barstow97a};
Wolff et al. \cite{Wolff98}; and references therein).

\begin{table}[t]
\caption[]{Identified interstellar metal lines in FUSE spectra of four
DA white dwarfs}
\begin{flushleft}
\begin{tabular}{lrcccc}
\noalign{\smallskip}
\hline
\noalign{\smallskip}
Ion      &\multicolumn{1}{c}{$\lambda$/\AA} 
                        &0455     &1631     &2309     &2331  \\
\noalign{\smallskip}
\hline
\noalign{\smallskip}
C\,II   &1036.34        &$\times$ &$\times$ &$\times$ &$\times$ \\
        &1037.02        &$\times$ &$\times$ &$\times$ &$\times$ \\
C\,III  &977.02         &$\times$ &$\times$ &$\times$ &$\times$ \\
N\,I    &952.30         &         &$\times$ &         &         \\
        &953.42         &         &$\times$ &$\times$ &$\times$ \\
        &953.66         &         &$\times$ &$\times$ &$\times$ \\
        &953.97         &         &$\times$ &$\times$ &$\times$ \\
        &954.10         &         &$\times$ &$\times$ &         \\
        &963.99         &         &$\times$ &$\times$ &$\times$ \\
        &964.63         &         &$\times$ &$\times$ &$\times$ \\
        &965.04         &         &$\times$ &$\times$ &         \\
        &1134.16        &$\times$ &$\times$ &$\times$ &$\times$ \\
        &1134.42        &$\times$ &$\times$ &$\times$ &$\times$ \\
        &1134.98        &$\times$ &$\times$ &$\times$ &$\times$ \\
N\,II   &915.61         &$\times$ &$\times$ &$\times$ &$\times$ \\
        &1083.99        &         &$\times$ &$\times$ &$\times$ \\
O\,I    &919.66         &         &$\times$ &$\times$ &$\times$ \\
        &921.86         &         &$\times$ &$\times$ &$\times$ \\
        &924.95         &         &$\times$ &$\times$ &$\times$ \\
        &929.52         &         &$\times$ &$\times$ &$\times$ \\
        &936.63         &         &$\times$ &$\times$ &$\times$ \\
        &948.69         &         &$\times$ &$\times$ &$\times$ \\
        &950.88         &         &$\times$ &$\times$ &$\times$ \\
        &971.74         &         &$\times$ &$\times$ &$\times$ \\
        &976.45         &         &$\times$ &$\times$ &$\times$ \\
        &988.66         &         &$\times$ &$\times$ &$\times$ \\
        &988.77         &$\times$ &$\times$ &$\times$ &$\times$ \\
        &1039.23        &$\times$ &$\times$ &$\times$ &$\times$ \\
Si\,II  &989.87         &$\times$ &$\times$ &$\times$ &$\times$ \\
        &1020.70        &         &$\times$ &$\times$ &$\times$ \\
P\,II   &963.80         &         &$\times$ &         &         \\
        &1152.82        &         &$\times$ &         &         \\
Ar\,I   &1048.22        &         &$\times$ &$\times$ &$\times$ \\
        &1066.66        &         &$\times$ &$\times$ &$\times$ \\
Fe\,II  &1063.18        &         &$\times$ &$\times$ &         \\
        &1096.88        &         &$\times$ &         &$\times$ \\
        &1121.98        &         &$\times$ &         &         \\
        &1125.45        &         &$\times$ &$\times$ &$\times$ \\
        &1144.94        &$\times$ &$\times$ &$\times$ &$\times$ \\

\noalign{\smallskip}
\hline
\noalign{\smallskip}
\label{ISM}
\end{tabular}
\end{flushleft}
\end{table}

\begin{table}[t]
\caption[]{Identified photospheric metal lines in FUSE spectra of three DA white
dwarfs}
\begin{flushleft}
\begin{tabular}{lr@{}lccc}
\noalign{\smallskip}
\hline
\noalign{\smallskip}
        &        & &0455 &2309 &2331  \\
Ion     &\multicolumn{1}{c}{$\lambda$/\AA} 
            &  &$W_{\lambda}$/m\AA\ &$W_{\lambda}$/m\AA\ &$W_{\lambda}$/m\AA\ \\
\noalign{\smallskip}
\hline
\noalign{\smallskip}
Si\,IV  &1066.61 &$^{\ast}$ \\
        &1066.65 &$^{\ast}$
                 &\raisebox{1.5ex}[-1.5ex]{71} &\raisebox{1.5ex}[-1.5ex]{30}
                 &\raisebox{1.5ex}[-1.5ex]{63} \\
        &1122.49 & &54   &24   &62 \\
        &1128.34 & &70   &38   &59 \\
P\,IV   &1030.52 & &10   &     &13 \\
P\,V    &1117.98 & &82   &36   &73 \\
        &1128.01 & &68   &37   &60 \\
S\,IV   &1062.68 & &     &     &11 \\
        &1073.00 & &16   &     &12 \\
S\,VI   &933.38  & &41   &     &21 \\
        &944.52  & &40   &8    &39 \\
\noalign{\smallskip}
\hline
\noalign{\smallskip}
\multicolumn{6}{l}{$^{\ast}$ possibly blended by interstellar Ar\,I
                   (1066.66~\AA)}\\
\label{PHOT}
\end{tabular}
\end{flushleft}
\end{table}

The four hot DA white dwarfs of our sample are
 -- with $T_{\rm eff} \ga 45\,000$~K -- in an interesting temperature region
for the study of radiative levitation. We have analyzed the FUSE spectra of
\object{GD\,246} (WD\,2309+105), \object{MCT\,0455$-$2812}
(WD\,0455$-$282, RE\,J0457$-$28),
\object{MCT\,2331$-$4731} (WD\,2331$-$475, RE\,J2334$-$47), and
\object{RE\,J1629+78} (WD\,1631+781). The most prominent features in all spectra
are the broad photospheric Lyman lines from L$\beta$ to the series limit.
In addition, sharp interstellar hydrogen features are always visible in the
line cores.
Further interstellar lines arise from several ground states of neutral or
low ionized heavier elements. Their Doppler velocities are compatible
with the hydrogen cores.
The strongest interstellar metal features are the N\,I lines near 1134~\AA,
C\,II at 1036.34~\AA, O\,I at 1039.23~\AA, Si\,II at 989.87~\AA, and
Fe\,II at 1144.94~\AA. A complete list can be found in Table~\ref{ISM}.

The spectra of MCT\,0455$-$2812, MCT\,2331$-$4731, and GD\,246 show a second
system of metal lines comprising several higher ionized species
(Si\,IV, P\,IV, P\,V, S\,IV, S\,VI) which are most probably of photospheric
origin. Their velocities 
are generally well-separated from the velocities of the interstellar lines.
Very prominent are the P\,V doublet at 1117.98/1128.01~\AA\ and the
Si\,IV doublet at 1122.49/1128.34~\AA. Sulfur could be detected in three objects
for the first time. Silicon lines have already been observed in IUE
spectra of GD\,246 (Holberg et al. \cite{Holberg98}). The ORFEUS observations
of MCT\,0455$-$2812 showed phosphorus and silicon
(Vennes et al. \cite{Vennes96}).
Si\,IV, P\,V, and S\,VI are typical ions which are also detected in FUSE
spectra of other hot DA white dwarfs (Chayer et al. \cite{Chayer2000};
Barstow et al. \cite{Barstow2001}).
The photospheric features found in our sample
are listed in Table~\ref{PHOT}. RE\,J1629+78 does not show photospheric metals.

\newpage
\subsubsection{Analysis}

Our main interests are the atmospheric element abundances
in order to improve our knowledge about the presence of heavier elements
and the radiative levitation process. For this purpose, we need to know
the structure of the atmosphere which is determined by the effective
temperature, the surface gravity, and the chemical composition. The composition
is extremely important due to the line-blanketing and backwarming effects
from millions of metal lines in the extreme ultraviolet (EUV). If this is not
taken into account the effective temperature from analyses of the Balmer lines
can be overestimated by several thousand degrees
(see e.g. Lanz et al. \cite{Lanz}; Wolff et al. \cite{Wolff98}).
The most important elements for the EUV opacity are iron and nickel.
The metals observed in our FUSE spectra are of minor importance and can
therefore be neglected for the atmospheric structure. However, we need
information about the iron and nickel content from other sources.

\begin{table*}[t]
\caption[]{Photospheric abundances relative to hydrogen (number ratios)
of the hot DA white dwarfs as determined
from FUSE spectra. Predicted values from Chayer et al. (\cite{Chayer95b})
are shown for comparison. For RE\,J1629+78, the calculation of Chayer et al.
with a pure hydrogen background is used. For the other stars, we list the
results calculated with an atmosphere contaminated with several heavier
elements}
\begin{flushleft}
\begin{tabular}{l@{}c@{}ll@{\,}lll@{\,}lll@{\,}lll@{\,}ll}
\noalign{\smallskip}
\hline
\noalign{\smallskip}
     &    &    &\multicolumn{3}{c}{Si/H} &\multicolumn{3}{c}{P/H}
               &\multicolumn{3}{c}{S/H}  &\multicolumn{3}{c}{Fe/H} \\
\multicolumn{3}{l}{WD Number} &\multicolumn{2}{l}{observed} &predicted
                              &\multicolumn{2}{l}{observed} &predicted
                              &\multicolumn{2}{l}{observed} &predicted
                              &\multicolumn{2}{l}{observed} &predicted \\
\noalign{\smallskip}
\hline
\noalign{\smallskip}
0455 &$-$ &282$^a$ & &$6.0\cdot10^{-7}$ &$2\cdot10^{-8}$
               &    &$1.2\cdot10^{-7}$ &n.a.
               &    &$1.0\cdot10^{-7}$ &$3\cdot10^{-6}$
               &$<$ &$1.0\cdot10^{-5}$ &$2\cdot10^{-6}$ \\
0455 &$-$ &282$^b$ & &$1.5\cdot10^{-6}$ &$1\cdot10^{-10}$
               &    &$2.0\cdot10^{-7}$ &n.a.
               &    &$5.0\cdot10^{-7}$ &$6\cdot10^{-6}$
               &$<$ &$1.0\cdot10^{-5}$ &$5\cdot10^{-6}$ \\
1631 &+   &781 &$<$ &$1.0\cdot10^{-9}$ &$2\cdot10^{-7}$
               &$<$ &$4.0\cdot10^{-10}$ & n.a.
               &$<$ &$3.0\cdot10^{-9}$ &$1\cdot10^{-6}$
               &$<$ &$2.0\cdot10^{-6}$ &$1\cdot10^{-6}$ \\
2309 &+   &105 &  &$5.0\cdot10^{-8}$  &$2\cdot10^{-8}$ 
               &  &$7.5\cdot10^{-9}$  &n.a.
               &$<$ &$3.0\cdot10^{-7}$ &$3\cdot10^{-6}$
               &$<$ &$2\cdot10^{-5}$   &$1\cdot10^{-6}$ \\
2331 &$-$ &475 &  &$6.0\cdot10^{-7}$  &$6\cdot10^{-9}$
               &  &$6.0\cdot10^{-8}$  &n.a.
               &  &$3.0\cdot10^{-7}$  &$2\cdot10^{-6}$
               &$<$ &$5\cdot10^{-6}$   &$5\cdot10^{-7}$ \\
\noalign{\smallskip}
\hline
\noalign{\smallskip}
\multicolumn{15}{l}{$^a$\,$T_{\rm eff}=56\,000$~K;
                    $^b$\,$T_{\rm eff}=66\,000$~K}\\
\label{DAres}
\end{tabular}
\end{flushleft}
\end{table*}

The four objects in our sample have been observed with the
Extreme Ultraviolet Explorer (EUVE). Wolff et al. (\cite{Wolff98})
have analyzed the EUVE data of white dwarfs using the well-studied DA
\object{G\,191-B2B} as reference object. They could reproduce the EUVE spectrum
with the same abundances as found from HST/GHRS spectra. We use their
EUV opacities for our objects to calculate model
atmospheres for the FUSE analysis.

Effective temperature and gravity could be determined in principle from the
Lyman lines of the FUSE spectra.
However, we will defer a complete detailed analysis of the broad hydrogen lines
until knowledge of the flux calibration and its uncertainties have been
improved, and use instead the following simpler procedure.
We use the results from Finley et al. (\cite{Finley})
for $\log g$ and an initial value of $T_{\rm eff}$,
calculate EUV-blanketed model atmospheres, and compare the
model line profiles to the observed FUSE spectra.
Since Finley et al. used pure hydrogen models, we have
to expect that they overestimated the effective temperatures. In this case, we
decrease $T_{\rm eff}$ until a reasonable fit to the FUSE observations can be
obtained. From the fit to different Lyman lines we estimate an uncertainty
of $\pm$\,2000~K for this procedure which is a sufficient accuracy for
the abundance analysis.

\begin{figure*}[htbp]
\centering
\includegraphics[width=\textwidth]{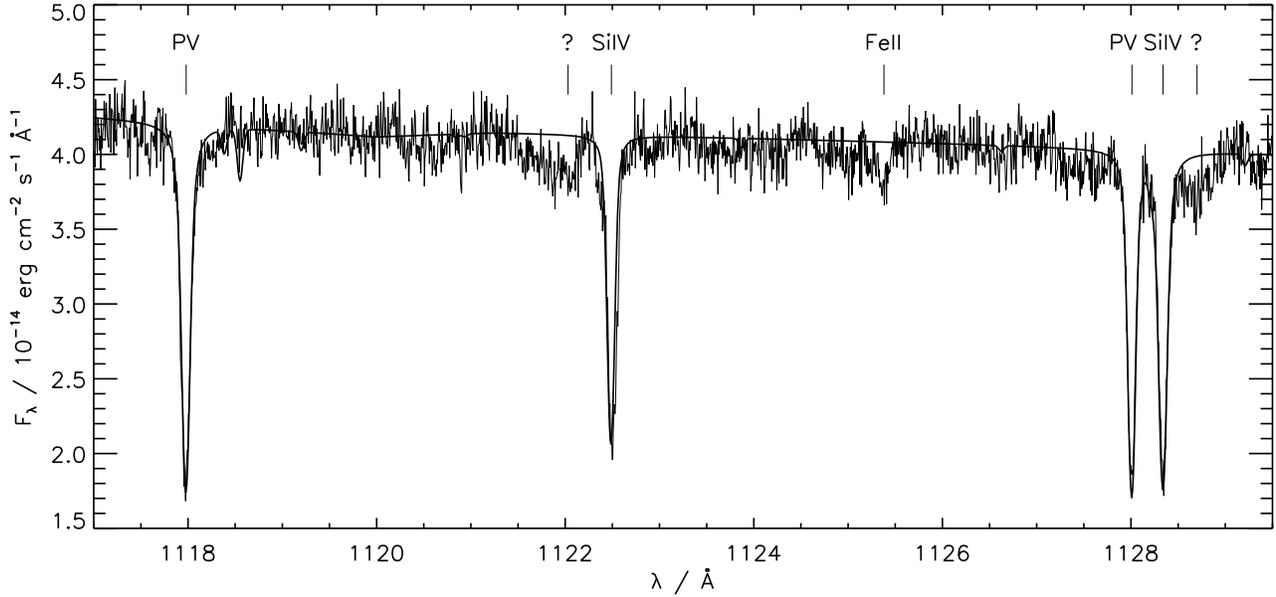}
\caption{Photospheric lines of Si\,IV and P\,V in the FUSE spectrum of
MCT\,2331$-$475 displayed together with a model spectrum. Also visible is an
interstellar line of Fe\,II (1125.45~\AA). The features at 1122.03~\AA\ and
1128.70~\AA\ are possibly due to S\,V and S\,IV, respectively. They are also
visible in MCT\,0455$-$2812 and
Feige\,55 and take part in the orbital motion of Feige\,55}
\label{figex1}
\end{figure*}

\begin{figure*}[htbp]
\centering
\includegraphics[width=\textwidth]{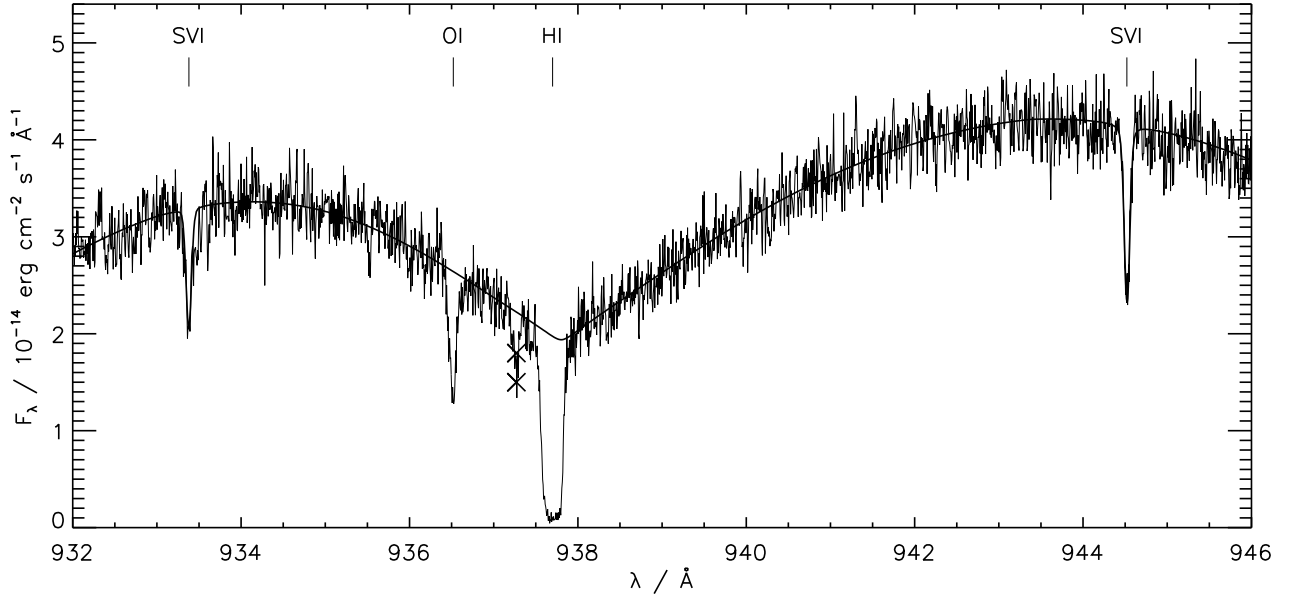}
\caption{Photospheric lines of S\,VI in the FUSE spectrum of
MCT\,2331$-$475 displayed together with a model spectrum. The feature at
937.3~\AA\ is an artifact}
\label{figex2}
\end{figure*}

We use line-blanketed LTE model atmospheres for the analysis of the Lyman
and metal lines. The program code is able to handle the absorption from
millions of metal lines in the EUV. A recent description of the procedures for
the calculation of theoretical atmospheres and synthetic spectra is given by
Finley et al. (\cite{Finley}).
For the abundance determination, we use a fixed atmospheric structure as
described above for each object. Synthetic spectra
for the FUSE region are then calculated with various abundances for
the interesting elements. The best fits are determined by a visual comparison
of observed and model spectra. Typical uncertainties in the abundances
caused by continuum fitting and the noise level are about 25\%.
The results are listed in Table~\ref{DAres}. Some example fits are presented
in Figs.~\ref{figex1} and \ref{figex2}. The resulting effective temperatures
are discussed in the next section.

\subsubsection{Comments on individual objects}

For GD\,246, Finley et al. (\cite{Finley}) have determined
$T_{\rm eff} = 58\,700$~K and $\log g = 7.8$. These values are compatible
with the FUSE Lyman lines if a pure hydrogen atmosphere is used.
With an atmosphere containing 25\%
of the G\,191-B2B abundances for the EUV absorption 
-- as implied by the EUVE analysis -- the Lyman series can be
reproduced better with $T_{\rm eff} = 54\,000$~K. We use this result
for the abundances listed in Table~\ref{DAres}. However, the values
do not change if a pure hydrogen atmosphere with $T_{\rm eff} = 59\,000$~K
is used instead. The photospheric lines in GD\,246 are at
$-6~{\rm km}\,{\rm s}^{-1}$ with respect to the interstellar lines,
the only case for which the separation is small.

Effective temperature and gravity for MCT\,2331$-$4731 from Finley et al.
are $55\,800$~K and 8.1, respectively.
These values are compatible with the FUSE spectrum, but
the abundance from the S\,VI lines is higher than the result from S\,IV.
However, both ionization stages can be reproduced with the same value if the
temperature is increased to 62\,000~K. We give this abundance in
Table~\ref{DAres}. The photospheric lines in MCT\,2331$-$4731 are at
$+27~{\rm km}\,{\rm s}^{-1}$ with respect to the interstellar lines.

The situation for MCT\,0455$-$2812 is more difficult.
Finley et al. have determined $T_{\rm eff} = 55\,700$~K
and $\log g = 7.8$ whereas Barstow et al. (\cite{Barstow97b}) found from
the ORFEUS Lyman spectrum $T_{\rm eff} = 66\,500$~K and $\log g = 7.4$.
The EUVE spectrum could also be reproduced better with
$T_{\rm eff} \approx 66\,000$~K (Dupuis et al. \cite{Dupuis95};
Barstow et al. \cite{Barstow97b}; Wolff et al. \cite{Wolff98}). 
Because we did not obtain sufficient data in the SiC channels, the FUSE data
cannot be used to distinguish between both temperatures. Therefore,
we list abundance determinations with both values in Table~\ref{DAres}.
As in the case of MCT\,2331$-$4731, the S\,IV and S\,VI lines cannot be
reproduced with the same abundance. The S\,VI value is about a factor of 10
higher even with $T_{\rm eff}=66\,000$~K. We give the S\,IV abundance
in Table~\ref{DAres}. In MCT\,0455$-$2812, the photospheric lines are at
$+64~{\rm km}\,{\rm s}^{-1}$ with respect to the interstellar lines.

The FUSE spectrum of RE\,J1629+78 does not show any photospheric
metal lines. This is in agreement with previous EUVE results
(Wolff et al. \cite{Wolff99a}, \cite{Wolff99b}) indicating a pure hydrogen
atmosphere. We list upper limits for Si, P, S, and Fe determined using
$T_{\rm eff}=42\,500$~K and $\log g = 7.6$ (Sion et al. \cite{Sion})
in Table~\ref{DAres}.

\subsection{The DAO Feige\,55}

\begin{figure*}[htbp]
\centering
\includegraphics[width=\textwidth]{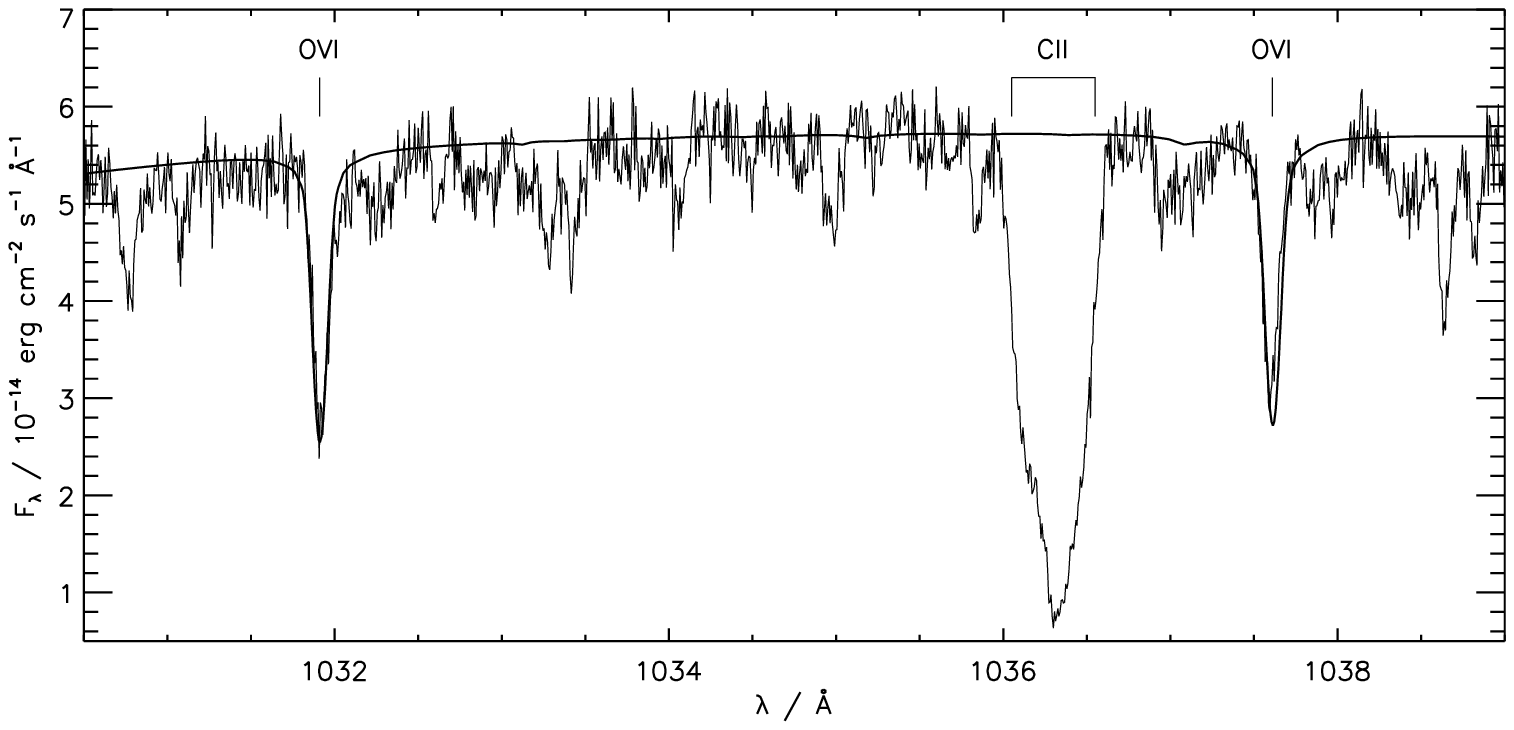}
\caption{O\,VI lines in the spectrum of Feige\,55 displayed together with
a model spectrum. The individual exposures were shifted according to the
orbital motion and co-added. Therefore, the interstellar C\,II line is
artificially broadened. There are also several other weak photospheric lines
present}
\label{figex3}
\end{figure*}

Feige\,55 (WD\,1202+608) is a double-degenerate white dwarf consisting of
a DAO and an unseen companion (Holberg et al. \cite{Holberg95b}).
Photospheric lines from carbon, nitrogen, oxygen, silicon,
iron, and nickel could be identified in IUE spectra
(Lamontagne et al. \cite{Lamontagne}, Holberg et al. \cite{Holberg98}).
The FUSE observations exhibit also numerous metal lines.
The photospheric features
can be easily distinguished from interstellar lines due to the orbital
motion which is visible in successive exposures. The new spectra add phosphorus
and sulfur to the list of identified metals in Feige\,55. Interesting is
the presence of the O\,VI lines at 1032/1038~\AA\ (see Fig.~\ref{figex3}).

At least two interstellar components can be resolved by FUSE, separated
by $45~{\rm km}\,{\rm s}^{-1}$. The blueward component has a much lower column
density and appears only in the strongest lines.
The velocities of the photospheric lines vary
from roughly $+13~{\rm km}\,{\rm s}^{-1}$ to $+90~{\rm km}\,{\rm s}^{-1}$,
relative to the stronger ISM component, during the P1042101 observation.
This is consistent with the semi-amplitude of $77~{\rm km}\,{\rm s}^{-1}$ and
the period of 1.5 days reported by Holberg et al. (\cite{Holberg95b}).

The determination of the effective temperature from Balmer lines is problematic
in Feige\,55 since the higher lines require higher temperatures than the lower
ones.
Bergeron et al. (\cite{Bergeron93}) could achieve a consistent fit to all lines
by adding iron with ${\rm Fe/H}=10^{-3}$ in addition to helium to the EUV
opacity. To avoid the Balmer line problem, Bergeron et al. (\cite{Bergeron94})
have used only the wings of the lines in their analysis of DAO white dwarfs.
They have determined
$T_{\rm eff}=58\,300 \pm 1500$~K, $\log g = 7.15 \pm 0.11$, 
$\log {\rm He/H} = -2.92 \pm 0.31$, and $T_{\rm eff}=55\,100 \pm 1300$~K,
$\log g = 6.96 \pm 0.10$, $\log {\rm He/H} = -2.91 \pm 0.28$, respectively,
from two optical spectra of Feige\,55.

For the following analysis,
we have fixed gravity and helium abundance at the values from
Bergeron et al. (\cite{Bergeron94}; $\log g =7.0$, ${\rm He/H}=10^{-3}$).
We do not use the He\,II lines of the FUSE spectra to determine the
helium abundance because these features are very broad and rather noisy
and do not allow a precise analysis. The value from Bergeron et al. is,
however, compatible with the FUSE lines.

$T_{\rm eff}$ is determined with the flux distribution from the optical to the
ultraviolet as given by the visual magnitude and the low resolution IUE
spectrum. As a first step, we have used only hydrogen and helium as absorbers.
With this result ($T_{\rm eff} = 65\,000$~K), preliminary abundances
for heavier elements were determined. In a second step, we added the most
important EUV absorbers iron and nickel to the opacity. This gave
$T_{\rm eff} = 60\,000 \pm 5000$~K.

The EUV opacity of helium, iron, and nickel is also taken into account for
the calculation of model spectra to determine the element abundances.
Two or more ionization stages are visible for most elements in the FUSE spectra.
We determine the element abundances seperately from each ionization stage
for three different effective temperatures. The results are shown in
Table~\ref{DAOres}. In addition to the ionization stages listed in this table,
there are also weak lines from O\,V and Fe\,VII present.

A systematic tendency can be observed in Table~\ref{DAOres}: 
At $T_{\rm eff} = 60\,000$~K, the higher ionization stage requires always
higher abundances than the lower stage. The discrepancy
can be removed if the temperature is increased to $70\,000$--$75\,000$~K.
These temperatures seem to be in contradiction with the overall flux
distribution. However, the gradient of the flux depends only weakly on
$T_{\rm eff}$ at these temperatures (Rayleigh-Jeans region) so that
errors of a few per cent in the IUE calibration or the visual magnitude
could result in changes of 10\,000~K. Moreover, the temperature determination
from Balmer lines is difficult, as described above, and the FUSE Lyman lines
favor also a somewhat higher temperature.
We cannot exclude that the different abundances are due to non-LTE effects
but the systematic tendency for all elements with more than one
ionization stage could indicate that the temperature is indeed higher.




\begin{table}[t]
\caption[]{Photospheric abundances (number ratios relative to hydrogen)
of Feige\,55 for different effective temperatures as derived from different
ionization stages}
\begin{flushleft}
\begin{tabular}{llll}
\noalign{\smallskip}
\hline
\noalign{\smallskip}
Ion    &$T_{\rm eff}=60\,000$\,K &$T_{\rm eff}=70\,000$\,K 
                                &$T_{\rm eff}=75\,000$\,K \\
\noalign{\smallskip}
\hline
\noalign{\smallskip}
C\,III &$1.0 \cdot 10^{-6}$     &$6.0 \cdot 10^{-6}$     &$2.0 \cdot 10^{-5}$ \\
C\,IV  &$4.5 \cdot 10^{-6}$     &$6.0 \cdot 10^{-6}$     &$1.0 \cdot 10^{-5}$ \\
N\,III &$1.0 \cdot 10^{-6}$     &$3.0 \cdot 10^{-6}$     &$3.0 \cdot 10^{-5}$ \\
N\,IV  &$2.0 \cdot 10^{-5}$     &$2.0 \cdot 10^{-5}$     &$3.0 \cdot 10^{-5}$ \\
O\,III &$4.0 \cdot 10^{-6}$     &$1.5 \cdot 10^{-5}$     &$3.0 \cdot 10^{-5}$ \\
O\,IV  &$6.0 \cdot 10^{-5}$     &$3.0 \cdot 10^{-5}$     &$3.0 \cdot 10^{-5}$ \\
O\,VI  &$3.0 \cdot 10^{-2}$     &$1.2 \cdot 10^{-4}$     &$3.0 \cdot 10^{-5}$ \\
Si\,IV &$6.0 \cdot 10^{-6}$     &$1.0 \cdot 10^{-5}$     &$1.5 \cdot 10^{-5}$ \\
P\,IV  &$1.0 \cdot 10^{-7}$     &$1.5 \cdot 10^{-7}$     &$4.0 \cdot 10^{-7}$ \\
P\,V   &$1.5 \cdot 10^{-7}$     &$1.5 \cdot 10^{-7}$     &$4.0 \cdot 10^{-7}$ \\
S\,IV  &$2.0 \cdot 10^{-7}$     &$3.0 \cdot 10^{-6}$     &$1.0 \cdot 10^{-5}$ \\
S\,VI  &$3.0 \cdot 10^{-6}$     &$3.0 \cdot 10^{-6}$     &$3.0 \cdot 10^{-6}$ \\
Fe\,V  &$3.0 \cdot 10^{-5}$     &$1.0 \cdot 10^{-4}$     &$1.0 \cdot 10^{-4}$ \\
Fe\,VI &$2.0 \cdot 10^{-4}$     &$1.0 \cdot 10^{-4}$     &$1.0 \cdot 10^{-4}$ \\
Ni\,V  &$5.0 \cdot 10^{-6}$     &$1.0 \cdot 10^{-5}$     &$1.0 \cdot 10^{-5}$ \\
\noalign{\smallskip}
\hline
\noalign{\smallskip}
\label{DAOres}
\end{tabular}
\end{flushleft}
\end{table}

\section{L$\beta$ satellites in CD\,$-38^{\circ}$\,10980}

\begin{figure*}[tbp]
\centering
\includegraphics[width=\textwidth]{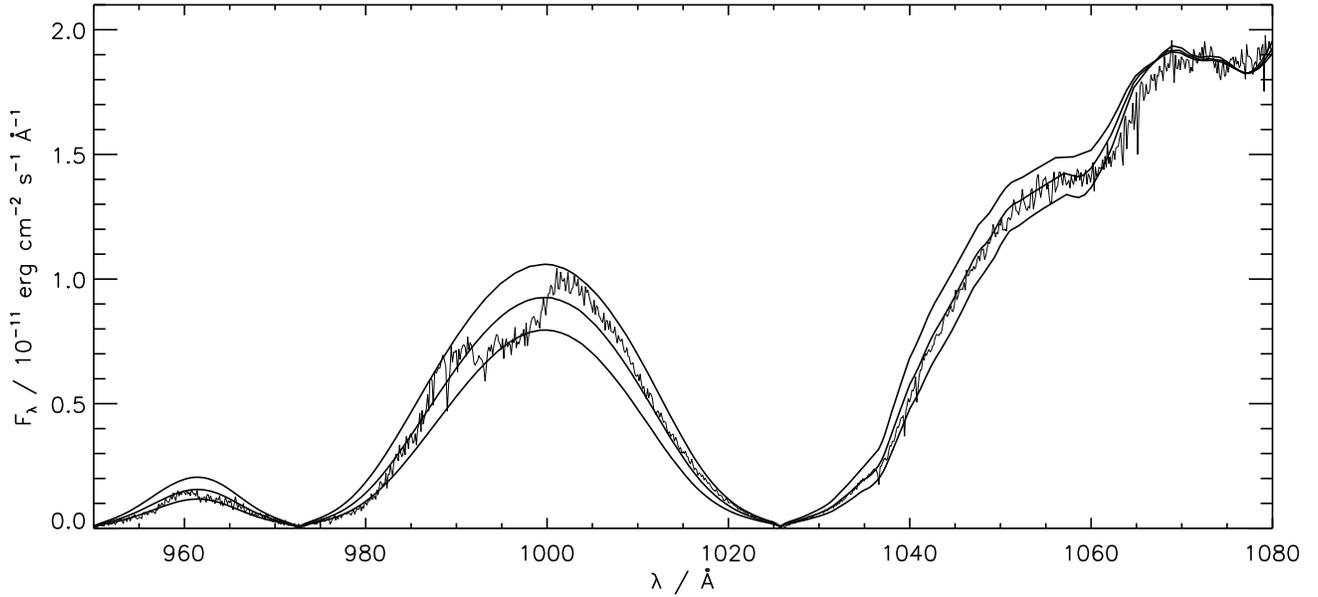}
\caption{FUSE spectra of CD\,$-38^{\circ}$\,10980, rebinned to 0.15~\AA\
resolution, and compared with three model
atmospheres for DA white dwarfs of 23\,000, 24\,000, 25\,000~K,
$\log g = 8.0$, normalized at 1070~\AA}
\label{figwd1620}
\end{figure*}

Cooler DA white dwarfs show peculiar absorption features in their
ultraviolet spectra, which have been identified as quasi-molecular
perturbation features of L$\alpha$ (Koester et al. \cite{Koester85};
Nelan \& Wegner \cite{NW}) or L$\beta$ (Koester et
al. \cite{Koester96}).  The features in the wing of L$\beta$ near 1060
and 1078~\AA\ are caused by a perturbation of the $n = 1 \rightarrow
3$ transition by protons; quantitatively the line profile can be
described by the theory developed over the last 20 years by Allard and
co-workers (e.g. Allard \& Kielkopf \cite{Allard91}; Allard \& Koester
\cite{Allard92}; Allard et al. \cite{Allard94}, Koester et
al. \cite{Koester96}; Allard et al. \cite{Allard98}).  A comparison
with observations of five DAs obtained with HUT and ORFEUS is
described by Koester et al. (\cite{Koester98}).

According to the theory, which involves the calculation of the
perturbed line profiles as well as the model atmosphere calculations
for DA white dwarfs, the satellite features of L$\beta$ should be
visible in the FUV in the temperature range of about 15000 to 25000~K.
In order to test this prediction, and also the accuracy of the
predicted shape of the satellite, we have observed
CD\,$-38^{\circ}$\,10980 (WD\,1620$-$391) with FUSE.  This object is
near the upper limit of the expected visibility range of the
satellites. Finley et al. (\cite{Finley}) obtained $T_{\rm eff} =
25\,280$~K, $\log g = 7.97$ from a fit to the optical Balmer lines;
Vauclair et al. (\cite{Vauclair97}), using a different set of optical
spectra but the same model atmospheres obtained 23\,230~K, $\log g =
8.13$, and used for their analysis of HIPPARCOS parallaxes the
averages of the two determinations: $T_{\rm eff} = 24\,250 \pm
1000$~K, $\log g = 8.05 \pm 0.10$.

Ten exposures of CD\,$-38^{\circ}$\,10980 were obtained on July 13,
2000. The LiF1 channel shows fluxes that are 
stable from one exposure to the next, so there are probably
little or no slit losses. We used the LiF1a exposures for our analysis which
cover the complete region of L$\beta$.
Fig.~\ref{figwd1620} shows the comparison of three model spectra
for $T_{\rm eff} = 23\,000$, 24\,000, 25\,000~K and $\log g = 8.0$
with the observed spectrum. For the region of the higher Lyman lines, we
added four of the exposures from SiC1b which show stable fluxes and are
consistent with LiF1a. The model fluxed have been fitted to the observation
near 1070~\AA.

The 24\,000~K model gives a fairly good fit to the observed line
profiles. The feature at 990 to 1000~\AA\ is a satellite feature of
L$\gamma$. It is visible in all channels and also in the HUT spectrum
of Wolf\,1346 (Koester et al. \cite{Koester96}) and in some ORFEUS spectra
(Koester et al. \cite{Koester98}). It is not yet included
in our models. In the 1100 to 1170~\AA\ region, which is not shown in the
figure, the fluxes are about 20\% lower than the models, which we
attribute to remaining absolute calibration uncertainties.  The
satellites are weak, but clearly visible also in this DA at about
24\,000~K. This confirms the expected disappearance of the features
near 25\,000~K. The 1060~\AA\ satellite is very well described by the
theoretical model atmosphere, which uses the line profile calculations
of Allard et al. (\cite{Allard98}) with the variable dipole moment
taken into account. The 1078~\AA\ satellite unfortunately is very
close to the edge of the channel and probably affected by some
calibration uncertainty, but does not seem to be in conflict
with theoretical prediction. The SiC2b exposure is better suited to study
this feature but suffers significantly from slit losses and is, therefore,
rather noisy. It seems, however, to confirm the theoretical prediction.

In addition to the Lyman series and the usual interstellar lines
also three lines from Si\,III can be observed in the FUSE spectra:
at 1108.358~\AA,
a blend at 1109.940, 1109.970~\AA, and a blend at 1113.204, 1113.230~\AA.
The longer-wavelength member of each blend is considerably stronger, so its
wavelength was used when computing the velocity. 
The Si\,III lines gave consistent results in each channel, and are 
at $+72 \pm 4~{\rm km}\,{\rm s}^{-1}$ with respect to the interstellar lines.

Holberg et al. (\cite{Holberg95a}) identified three different
velocity systems in IUE echelle spectra of CD\,$-38^{\circ}$\,10980:
interstellar lines with
$V_{\rm ISM}=-30.10 \pm 1.03\,{\rm km}\,{\rm s}^{-1}$,
circumstellar lines with $V_{\rm CSM}=+39.25 \pm 0.94\,{\rm km}\,{\rm s}^{-1}$,
and Si\,IV lines (1393/1402~\AA) with
$V_{\rm Si\,IV}=+46.6 \pm 3.2\,{\rm km}\,{\rm s}^{-1}$. The photospheric
velocity is $V_{\rm phot}=+51.4 \pm 2.0\,{\rm km}\,{\rm s}^{-1}$.
The velocity of the FUSE Si\,III lines is halfway between
the circumstellar and photospheric components and is consistent with the
Si\,IV lines.

The Si\,III lines can be reproduced using a model atmosphere of
$T_{\rm eff}=24\,250$~K and $\log {\rm Si/H} = 2 \cdot 10^{-8}$.
The same abundance
can also account for the equivalent widths -- as determined by 
Holberg et al. (\cite{Holberg95a}) -- of the Si\,IV and the
circumstellar Si\,III and Si\,II lines. The success of a single model
makes a common photospheric origin of all lines plausible, although we
do not have an explanation for the discrepant velocities.

\section{Discussion}

The different amounts of metals found in the FUSE spectra of the four
hot DA white dwarfs correspond to the observed opacity in the extreme
ultraviolet (see Wolff et al. \cite{Wolff98}, \cite{Wolff99a}, \cite{Wolff99b}).
The two objects with the strongest EUV absorption
(MCT\,0455$-$2812 and MCT\,2331$-$4731) have the largest FUSE abundances
whereas GD\,246 shows considerably lower values. Both the EUVE and the FUSE
spectra of RE\,J1629+78 do not exhibit signs of additional photospheric
absorption. These results show that the EUV is in general a good indicator
for the total amount of metals although this spectral region is mainly
sensitive to the abundances of iron and nickel. A common
reason for the presence of iron and other heavier elements seems therefore
to be plausible. The iron abundances implied by the EUVE analysis
($5\cdot10^{-6}$ or lower) are compatible with the upper limits from the
FUSE spectra. 

The generally accepted explanation for heavier elements is radiative levitation.
In Tables~\ref{DAres} and \ref{DAOpre}, we list theoretical predictions from
Chayer et al. (\cite{Chayer95b}) for the hot DAs and for Feige\,55.
We use the results calculated with an atmosphere contaminated by several
heavier elements for the three hottest DAs.
Such an atmosphere should be the best approximation to the real situation.
For Feige\,55, the calculations with a pure hydrogen background plasma
must be taken because models with a mixed hydrogen-helium atmosphere are not
available and the contaminated models do not extend to
$T_{\rm eff} = 70\,000$~K and higher.

The comparison with the observations shows the well known result that predicted
and measured abundances agree rather poorly for individual elements
(see e.g. Chayer et al. \cite{Chayer95a}). The observed silicon abundances
are higher than the predictions with the exception of the coolest DA
RE\,J1629+78.
A similar dependence on the effective temperature can be observed in other
DAs as well (see Holberg et al. \cite{Holberg97} for a discussion).
Sulfur has always lower measured abundances whereas the iron abundances are
compatible with the predictions.

\begin{table}[t]
\caption[]{Predicted abundances of the radiative levitation theory for
Feige\,55. 
We list the results by Chayer et al. (\cite{Chayer95b}) calculated with a pure
hydrogen background plasma}
\begin{flushleft}
\begin{tabular}{llll}
\noalign{\smallskip}
\hline
\noalign{\smallskip}
El. &$T_{\rm eff}=60\,000$\,K &$T_{\rm eff}=70\,000$\,K 
                              &$T_{\rm eff}=75\,000$\,K \\
\noalign{\smallskip}
\hline
\noalign{\smallskip}
C   &$1 \cdot 10^{-5}$        &$1 \cdot 10^{-5}$        &$8 \cdot 10^{-6}$ \\
N   &$2 \cdot 10^{-5}$        &$4 \cdot 10^{-5}$        &$5 \cdot 10^{-5}$ \\
O   &$2 \cdot 10^{-5}$        &$5 \cdot 10^{-5}$        &$6 \cdot 10^{-5}$ \\
Si  &$1 \cdot 10^{-7}$        &$6 \cdot 10^{-8}$        &$5 \cdot 10^{-8}$ \\
S   &$2 \cdot 10^{-5}$        &$3 \cdot 10^{-5}$        &$3 \cdot 10^{-5}$ \\
Fe  &$4 \cdot 10^{-5}$        &$8 \cdot 10^{-5}$        &$1 \cdot 10^{-4}$ \\
\noalign{\smallskip}
\hline
\noalign{\smallskip}
\label{DAOpre}
\end{tabular}
\end{flushleft}
\end{table}

The discrepancies between observed and predicted abundances in hot white dwarf
atmospheres are probably due to the limitations of previous calculations.
Some of these inaccuracies could be removed by Dreizler \& Wolff (\cite{DW})
who considered for the first time the influence of the absorbing ions on the
atmospheric structure self-consistently. The only free parameters of their
models are $T_{\rm eff}$ and $\log g$ -- the element stratification and the
model spectra are determined directly by these parameters. While this model
could reproduce the EUVE spectrum of G\,191-B2B -- which is mainly determined
by iron and nickel -- better than atmospheres without chemical stratification,
the ultraviolet lines of carbon, nitrogen, and silicon
could not be reproduced with similar accuracy. This is probably due to
remaining uncertainties of the model calculations.

Another problem is the observed helium abundance in Feige\,55 which is also
difficult to understand with radiative levitation.
The predicted abundance is lower than the observed value
(Vennes et al. \cite{Vennes88}) and a stratified hydrogen-helium atmosphere
is in contradiction with the observed optical helium lines
(Bergeron et al. \cite{Bergeron94}). Theoretical calculations by
Unglaub \& Bues (\cite{UB98}) have shown that mass loss may play an important
role for the presence of helium in DAO white dwarfs and the transition of DOs
into DAs. Unglaub \& Bues (\cite{UB2000}) predict for a star with the
parameters of Feige\,55 ($T_{\rm eff} \approx 60\,000$~K, $\log g \approx 7$)
a helium abundance of ${\rm He/H} \ga 10^{-3}$. This correponds well to the
measured value. If mass loss is indeed important in DAOs then it cannot be
expected that the abundances of heavier elements follow the predictions of
equilibrium calculations. This could explain the failure of the radiative
levitation theory.

The DA white dwarfs in our sample have significantly higher gravities than
Feige\,55. In these objects, mass loss is probably not important. Therefore,
improved equilibrium calculations are necessary to test the equilibrium
radiative levitation theory.

In conclusion, we can say that the unique capabilities of FUSE complement
other ultraviolet observations for studying
heavy elements in hot white dwarfs. FUSE gives access to
otherwise undetectable elements and provides new details
in understanding the roles of radiative levitation and mass loss.

\begin{acknowledgements}
This work is based on data obtained for the Guaranteed Time Team by the
NASA-CNES-CSA FUSE mission operated by the Johns Hopkins University.
Financial support to U.S. participants has been provided by NASA contract
NAS5-32985. BW acknowledges support by the Deutsches Zentrum
f\"ur Luft- und Raumfahrt (DLR) under grant 50 OR 96173.
\end{acknowledgements}

\listofobjects

\end{document}